\documentclass[pra,floatfix,10pt]{revtex4}

\usepackage{amsmath}
\usepackage{amsfonts}
\usepackage{graphicx}

\begin{document}

\author{Lian-Ao Wu}
\affiliation{Chemical Physics Theory Group, and Center for Quantum
  Information and Quantum Control, Chemistry Department, University of Toronto,
80 St. George St., Toronto, Ontario M5S 3H6, Canada}
\author{Daniel A. Lidar}
\affiliation{Chemical Physics Theory Group, and Center for Quantum
  Information and Quantum Control, Chemistry Department, University of Toronto,
80 St. George St., Toronto, Ontario M5S 3H6, Canada}
\title{Overcoming Quantum Noise in Optical Fibers}

\begin{abstract}
Noise in optical Telecom fibers is an important limitation on optical
quantum data transmission. Unfortunately, the classically successful
amplifiers (such as EDFA) cannot be used in quantum communication because of
the no-cloning theorem. We propose a simple method to reduce quantum noise:
the insertion of phase-shifters and/or beam-splitters at regular distance
intervals into a fiber. We analyze in detail the case of qubits encoded into
polarization states of low-intensity light, which is of central importance
to various quantum information tasks, such as quantum cryptography and
communication. We discuss the experimental feasibility of our scheme and
propose a simple experiment to test our method.
\end{abstract}

\maketitle

\section{Introduction}

Quantum communication (QC) has recently emerged as a subject of much
interest, due to its applications in distributed quantum computation and
quantum cryptography \cite{Bouwmeester:book}. In QC, non-orthogonal quantum
signals are typically transmitted through Telecom fibers. Reducing noise in
Telecom fibers is crucial for QC applications, because the very weak signals
carried by polarization states are usually employed. Ideally, a single
photon in a superposition of two pulses separated in time with a controlled
phase difference (i.e., $|\mathrm{pulse}~A\rangle +e^{i\theta }|\mathrm{pulse}~B\rangle $) may be used. However, on-demand single-photon sources remain
an important technological challenge. Currently, weak coherent states are
often employed as approximate single photon signals. It is well known in
quantum cryptography \cite{Gottesman:00,Gisin:02}) -- a branch of QC -- that weak
coherent states may open up loopholes in security because of the
probabilistic existence of multi-photon signals \cite{Lutkenhaus:99}. Indeed, a weak
coherent state, when phase randomized via decoherence, gives a Poisson
distribution in photon numbers. An eavesdropper, Eve, may, in principle,
measure the photon number in such a signal by a non-demolition measurement.
Afterward, she can stop all single photon signals from reaching the
receiver, Bob. For each multi-photon signal, she can steal one photon and
keep it in her quantum memory and send the rest of the signals to Bob by
using, for example, an ideal channel instead. Since Eve now has an exact
copy of the quantum state transmitted to Bob, this creates a significant
challenge in the security of quantum cryptography \cite{Lutkenhaus:99}. Thus,
attenuation losses and decoherence in QC are a major issue and methods for
reducing such quantum noise are therefore important. Unfortunately, the
classically successful amplifiers (such as EDFA \cite{Newell:book}) cannot
be used in QC because of the no-cloning theorem \cite{Zurek}, and new
methods must be explored.

Here, we propose a method to reduce noise in the transmission of quantum
optical signals in a Telecom fiber. Our method is inspired by the theory of
quantum dynamical \textquotedblleft bang-bang\textquotedblright\ (BB)
decoupling \cite{Viola:98}. However, a key novelty of our work is the
following: we propose to implement BB control in space, rather than time,
through the insertion at regular intervals of a sequence of simple linear
optical elements (phase-shifters and/or beam-splitters) in sections of a
Telecom fiber. We also discuss the experimental feasibility of our scheme,
and propose a few simple experimental tests. We do not expect our method to
improve the fidelity of classical light transmission compared to, e.g., EDFA
amplifiers, since our method turns out to be quite sensitive to reflection
from optical elements and deviations from average fiber homogeneity, which
is not the case for classical amplifiers.

\section{Quantum noise in optical fibers}

\label{noise}

An optical fiber provides boundary conditions that guide light along a
(locally) straight trajectory. An ideal fiber allows modes of traveling
photons to propagate through unchanged. A real fiber induces noise
(dispersion, loss, decoherence) compared to the ideal case. The method we
introduce in this paper is designed to cancel (in principle) all quantum
noise. The dominant classical-light loss mechanisms in an optical fiber are
UV absorption, Rayleigh backscattering, OH absorption, and Raman scattering,
and lead to typical attenuation rates, for state-of-the-art commercial
silica Telecom fiber, of about 0.25db/km \cite{Newell:book}. These
mechanisms are active also in the quantum regime \cite{Drummond:99,Drummond:99a}. All
noise processes affecting quantum light in optical fibers are derivable from
a microscopic Hamiltonian describing (i) the direct interaction between
photons and the optical (dielectric) material of a fiber, and (ii) the
indirect interaction between photons and quasi-particle excitations of the
fiber material, such as polaritons and photon-phonon coupling. These
indirect interactions are, of course, in turn derivable from a microscopic
Hamiltonian that takes into account matter-matter interactions in the fiber,
and couples them to photons. The derivation of the resulting effective
interactions (e.g., a non-linear Schr\"{o}dinger equation) from such
microscopic Hamiltonians has been covered in detail, e.g., in \cite{Drummond:99,Drummond:99a,Mitchell:00}.

The starting point of our analysis is the observation that all interactions
involving photons can be written in terms of polynomials in the bosonic
raising and lowering operators $b_{j}^{\dagger },b_{j}$ (where $j$ is the
mode of the traveling photons \cite{Drummond:99,Drummond:99a}). A polynomial of order $N$
describes an $N$-photon process, and typically the cross-section of
interactions decreases with increasing $N$. In the case of the
non-relativistic quantum electrodynamics of charged particles one can
decompose the photon-matter interaction Hamiltonian into linear and
quadratic terms with respect to the photon field, $H_{I}=H_{I}^{l}+H_{I}^{q}$%
, where the linear part is 
\begin{equation}
H_{I}^{l}=\sum_{j}(b_{j}\hat{B}_{j}^{\dagger }+b_{j}^{\dagger }\hat{B}_{j}),
\label{eq2}
\end{equation}%
where the \textquotedblleft bath\textquotedblright\ operators $\hat{B}_{j}$
depend only on the variables of charged particles and/or quasi-particles,
and the quadratic part $H_{I}^{q}$ is a function of the bilinear operators, $%
b_{i}^{\dagger }b_{j}$, $b_{i}^{\dagger }b_{j}^{\dagger }$ and $b_{i}b_{j}$.
Higher-order interactions may originate from relativistic effects. In
general $H_{I}^{q}$, which makes no contribution to one-photon processes, is
much smaller than $H_{I}^{l}$ \cite{Cohen-Tannoudji}. Therefore, the
quadratic term can usually be neglected.

Let us substantiate these arguments by briefly reviewing the corresponding
non-relativistic electrodynamics. Consider particles $\alpha $ with charge $%
q_{\alpha }$ and mass $m_{\alpha }$, which constitute the optical material
of a fiber. Let $\mathbf{r}_{\alpha }$ and $\mathbf{p}_{\alpha }$ be the
position and momentum of particle $\alpha $, and $\mathbf{A}(\mathbf{r})$ be
the vector potential of the photon field. The system-bath Hamiltonian that
describes the dynamics can be written, in the Coulomb gauge, as 
\begin{eqnarray}
H &=&H_{0}+H_{I};  \notag \\
H_{0} &=&H_{M}+H_{P}  \label{eq:H0}
\end{eqnarray}%
Here $H_{M}$ depends only on the variables of the charged particles. $%
H_{P}=\sum \hbar \omega _{j}(b_{j}^{\dagger }b_{j}+1/2)$ is the free photon
Hamiltonian, where $b_{j},b_{j}^{\dagger }$ are the photon annihilation and
creation operators in the normal vibrational mode $j$ of the field
identified by the wave vector $\mathbf{k}_{j}$, the polarization $\mathbf{%
\varepsilon }_{j}$ and the frequency $\omega _{j}=ck_{j}$, where $c$ is the
speed of light in vacuum. Then the linear part with respect to the photon
field \cite{Cohen-Tannoudji} is 
\begin{eqnarray}
H_{I}^{l} &=&\sum_{\alpha }(\frac{q_{\alpha }}{m_{\alpha }}\mathbf{p}%
_{\alpha }\cdot \mathbf{A}(\mathbf{r}_{\alpha })+\frac{g_{\alpha }q_{\alpha }%
}{2m_{\alpha }}\mathbf{S}_{\alpha }\cdot \mathbf{B}(\mathbf{r}_{\alpha }))
\label{eq:Hl} \\
&=&\sum_{j}(\hat{B}_{j}^{\dagger }b_{j}+\hat{B}_{j}b_{j}^{\dagger })  \notag
\end{eqnarray}%
where for a cubic box with dimension $L$ the operator $\hat{B}_{j}$ can be
expressed as 
\begin{equation*}
\hat{B}_{j}^{\dagger }=-\sum_{\alpha }\frac{q_{\alpha }}{m_{\alpha }}\sqrt{%
\frac{\hbar \omega _{j}}{2\varepsilon _{0}L^{3}}}e^{i\mathbf{k}_{j}\cdot 
\mathbf{r}_{\alpha }}(\mathbf{p}_{\alpha }\cdot \mathbf{\varepsilon }_{j}+%
\frac{ig_{\alpha }}{2c}\mathbf{S}_{\alpha }\cdot \mathbf{k}_{j}\times 
\mathbf{\varepsilon }_{j}),
\end{equation*}%
which only depends on the variables of charged particles. Here $g_{\alpha }$
is the $g$ factor, $\varepsilon _{0}$ is the permittivity of free space, and 
$\mathbf{S}_{\alpha }$ is the spin of particle $\alpha $. Note that the
interaction is linear in the operators $b_{j}$ and $b_{j}^{\dagger }$.

The quadratic part of the interaction Hamiltonian is found to be 
\begin{equation}
H_{I}^{q}=-\sum_{\alpha }\frac{q_{\alpha }^{2}}{2m_{\alpha }}\mathbf{A}^{2}(%
\mathbf{r}_{\alpha }),  \label{eq:q}
\end{equation}%
and is a function of the bilinear operators, $b_{i}^{\dagger }b_{j}$, $%
b_{i}^{\dagger }b_{j}^{\dagger }$ and $b_{i}b_{j}$ .

Under the long-wavelength approximation, where the spatial variations of the
electromagnetic field over the size of the particles is negligible, first
order perturbation theory of $H_{I}^{l}$ results in the widely applied
dipole interaction (e.g., \cite{Mitchell:00} and references therein). Some
effective interactions, such as atom-mediated photon-photon interactions and
nonlinear photon-photon interactions (Kerr effect) have been derived without
consideration of $H_{I}^{q}$ \cite{Drummond:99,Drummond:99a,Mitchell:00}. We
provide more details on these effective interaction in Section \ref{leading}%
. For simplicity of presentation we will first design an \textquotedblleft
anti-\emph{linear}-decoherence fiber\textquotedblright\ by considering $%
H_{I}^{l}$ only. Later on we show how to treat higher order interaction
terms. It is important to stress that in essence our method hardly depends
on the details of the interaction, but \emph{depends on the statistics of
photons as bosons}. For this reason our method is very general and is in
principle applicable to the entire phenomenology of quantum noise processes
affecting photons in fibers, though its practical applicability is a matter
of being able to satisfy certain constraints that will be discussed in
detail below.

\section{The Anti-Linear-Decoherence Fiber}

We first consider quantum data transmission through a Telecom fiber with
noise induced by $H_{I}^{l}$. Since $H_{I}^{l}$ describes absorption and
creation of photons, it generates photon loss, among other processes. To
simplify, we suppose that a \emph{polarization} photon is transmitted from
end $A$ to end $B$. One can define a logical qubit supported by $\left\vert
0\right\rangle _{L}=b_{1}^{\dagger }\left\vert \mathrm{vac}\right\rangle $
and $\left\vert 1\right\rangle _{L}=b_{2}^{\dagger }\left\vert \mathrm{vac}%
\right\rangle $ where the mode indices refer to the two polarization states.
The initial state at end $A$ is $\left\vert \Psi _{A}\right\rangle
=(a\left\vert 0\right\rangle _{L}+b\left\vert 1\right\rangle _{L})\left\vert
M\right\rangle $, where $\left\vert M\right\rangle $ is the state of the
bath (dielectric material and quasi-particle excitations in the fiber). At
the time $T=X/v$ (where $X$ is the distance between $A$ and $B$, and $v$ is
the average speed of light in the Telecom fiber) the wave function is $%
\left\vert \Psi (T)\right\rangle =U(T,0)$ $\left\vert \Psi _{A}\right\rangle 
$, where the evolution operator is (in units where $\hbar =1$) $%
U(T,0)\approx e^{-iH(N\Delta )\tau }\cdots e^{-iH(2\Delta )\tau
}e^{-iH(\Delta )\tau }$, where $H(k\Delta )\equiv \frac{1}{\Delta }%
\int_{(k-1)\Delta }^{k\Delta }[H_{I}(x)+H_{0}(x)]dx$ is the average
Hamiltonian over the $k$th segment, where $H_{0}$ is a sum of the matter
(and/or excitations) and photon self-Hamiltonians, $\tau =\Delta /v$, and we
have assumed that $N=X/\Delta $ is large in order to expand the
normal-ordered exact propagator $U(T,0)=\,:\exp
[-i\int_{A}^{B}[H_{I}(x)+H_{0}(x)]dx]:$. I.e., we have neglected deviations
from average fiber homogeneity, $\delta _{k}=\langle (H(k\Delta
)-[H_{I}(k\Delta )+H_{0}(k\Delta )])^{2}\rangle $ [$U(T,0)$ can easily be
expressed including such second and higher order moments using a Magnus
expansion, and it is known how to generalize BB decoupling to treat such
higher moments, at the expense of more BB pulses \cite{Viola:99}]. The
interaction $H_{I}$ entangles the output wave function at end $B$ with the
material/excitations in the fiber. By standard arguments it follows that,
therefore, the quantum information encoded into the photon state will
decohere \cite{Bouwmeester:book}.

In order to solve this problem of decoherence, we draw inspiration from the
idea of BB decoupling via time-dependent pulses \cite{Viola:98} (we note
that a method for finding such pulses directly from empirical data was
proposed in \cite{ByrdLidar:02}). We first recall the action of a
phase-shifter. It is simple to show [using the Baker-Campbell-Hausdorff
(BCH) formula \cite{Reinsch:00}] for a boson that 
\begin{equation}
e^{i\phi \hat{n}}b^{\dagger }e^{-i\phi \hat{n}}=e^{i\phi }b^{\dagger },\quad
e^{i\phi \hat{n}}be^{-i\phi \hat{n}}=e^{-i\phi }b,  \label{eq:n-b}
\end{equation}%
where $\hat{n}=b^{\dagger }b$ is a boson number operator. Physically, the
operation $e^{i\pi \hat{n}}$ is a $\pi $ phase-shifter (it puts a phase of $%
\pi $ between the number states $|0\rangle $ and $|1\rangle $, not to be
confused with our logical qubit states). Defining the $\pi $-phase-shifter
operator 
\begin{equation}
\Pi =\Pi ^{\dagger }=e^{i\pi (\hat{n}_{1}+\hat{n}_{2})},
\end{equation}%
we therefore have 
\begin{equation}
\Pi H\Pi =H_{0}-H_{I}^{l},  \label{eq:flip}
\end{equation}%
because the photons term of $H_{0}$ is $\sum \hbar \omega _{j}(n_{j}+1/2)$,
so that $[H_{0},n_{1}+n_{2}]=0$. The crucial point is that \emph{the sign of
the linear term of the interaction Hamiltonian has been negated by the
action of two phase-shifters, i.e., effectively time-reversed}. Now, if we
install thin phase-shifters inside the fiber at positions $x=0,\Delta
,2\Delta ,...$, from $A$ to $B$, the evolution will be modified to 
\begin{eqnarray*}
U^{\prime }(T,0) &\approx &e^{-iH(N\Delta )\tau }\cdots \Pi e^{-iH(2\Delta
)\tau }\Pi e^{-iH(\Delta )\tau }\Pi \\
&\equiv &[N,...,\Pi ,2,\Pi ,1,\Pi ],
\end{eqnarray*}%
where in the second line we have introduced a self-explanatory notation that
will be used repeatedly below. Note that in writing this expression we have
neglected the variation of $H$ inside the phase-shifter; this will hold
provided that the phase-shifter width is much smaller than the distance over
which deviations $\delta _{k}$ from average fiber homogeneity become
significant. Further note that we are applying the \textquotedblleft
parity-kick\textquotedblright\ version of BB decoupling \cite%
{Viola:98,Vitali:99}, but are implementing it in space, rather than time.
Now assume that the \emph{average} Hamiltonians over two successive segments
are equal: 
\begin{eqnarray}
H_{I}^{l}((k+1)\Delta ) &=&H_{I}^{l}(k\Delta )  \notag \\
H_{0}((k+1)\Delta ) &=&H_{0}(k\Delta ).  \label{eq:approxH}
\end{eqnarray}%
The better this approximation, the better our method will perform; we
address deviations in Appendix~\ref{app}. In this case, to first order in $%
\tau $, and using Eq.~(\ref{eq:flip}), we have exact cancellation of $%
H_{I}^{l}$ between successive segments: 
\begin{eqnarray}
e^{-iH((k+1)\Delta )\tau }\Pi e^{-iH(k\Delta )\tau }\Pi =
e^{-iH((k+1)\Delta )\tau }e^{-i\Pi H(k\Delta )\Pi \tau }
= e^{-2iH_{0}(k\Delta )\tau }.  \label{eq:cancel}
\end{eqnarray}%
This yields the overall evolution operator 
\begin{equation*}
U^{\prime }(T,0)=e^{-iH_{0}(X)N\tau }=e^{-iH_{0}(X)T},
\end{equation*}%
i.e., the evolution is completely decoherence-free, in analogy to the ideal
BB limit of infinitely fast and strong pulses \cite{Viola:98}.

\section{Rough estimate of required inter-phase-shifter distance}

Because of the in-principle equivalence between the BB method and the quantum
Zeno effect \cite{Facchi:03}, the proposed method can only work if the
phase-shifters are inserted at small intervals $\Delta $ over which
coherence loss is quadratic (\textquotedblleft Zeno-like\textquotedblright
), rather than exponential (\textquotedblleft Markovian\textquotedblright ).
A reliable estimate of $\Delta $ requires a first principles calculation
which is beyond the scope of the present work; we present a phenomenological
model for a detailed estimate of $\Delta $ in Appendix~\ref{app}. Here we
give a rough \emph{upper bound }estimate of this distance. We assume that
the linear term of the interaction Hamiltonian gives rise to the $0.25$dB/km
($5\times 10^{-2}$) \emph{classical} loss figure in a Telecom fiber. Our
main approximation now consists in further assuming that the insertion of
phase-shifters into the fiber causes a reduction of loss from first to
second order, and we use this to estimate the $\Delta $ required in the 
\emph{quantum} case. Thus, imagine a distributed quantum computing scenario
where small-scale quantum computers are connected by optical fibers of
length about 1km. Our goal is to have reliable quantum computation within
the fault-tolerance threshold value of $10^{-4}$ error rate for each
elementary quantum logical operation. [We remark that for reliable quantum 
\emph{communication of entangled photon pairs}, the current error rate of
about $5\times 10^{-2}$ is already acceptable provided one allows the
application of entanglement purification \cite{Dur:99}; our scheme is
significantly simpler.] Therefore, we need to cut down the loss figure from $%
5\times 10^{-2}$ to say $10^{-4}$. Suppose we need to insert $N$
phase-shifter within 1km of a Telecom fiber. Denote the attenuation between
a pair of phase-shifter by $l$. Then, without the $N$ phase-shifters, we
have $(1-l)^{N}=0.95$. For a sufficiently large $N$, we can expand the
expression binomially and obtain the approximation $lN=0.05$. Now, with the
insertion of phase-shifters, we simply assume that the attenuation between
two phase-shifters is due to a second order contribution of the form $l^{2}$%
. We further assume that those contributions sum up in usual addition.
Therefore, we have $l^{2}N=10^{-4}$. This yields $l=2\times 10^{-3}$ and $%
N=25$. Recalling that two phase-shifters are needed per cancellation step,
we see that about $50$ phase-shifters have to be inserted in a distance of
1km which translates to one phase-shifter every $20$m. This figure is merely
a rough upper bound estimate on the distance $\Delta $ between two
phase-shifters for our scheme to be useful; one can also determine $\Delta $
via the experiment we propose below. Also note that we have assumed here
that the fiber is straight as is typically done in theoretical models. In
order to regain the straight fiber approximation, in the case of a curved
fiber $\Delta $ is upper-bounded by the local radius of curvature.

While in spirit our method is similar to BB decoupling \cite{Viola:98}, a
major advantage here is that we do not need to apply any time-dependent
pulses, which may result in significant uncertainties such as gate errors
and off-resonance transitions. Instead, the phase-shifters may be
incorporated into the fiber directly during the manufacturing process.
Alternatively, time-independent (say, electronic or pressure) controls may
be applied at various points of a Telecom fiber to achieve the action of
pulse shifters.

\section{The Anti-Bilinear-Decoherence Fiber}

We now consider higher order processes. Although they are generally weak,
the bilinear interactions appearing in $H_{I}^{q}$ may still cause
decoherence. A direct harmful consequence is to change the polarization
direction, through a term such as $b_{1}^{\dagger }b_{2}$. In the classical
case, the fiber structure can be designed so that a\emph{\ known }
polarization direction can be preserved \cite{Agrawal:book}. In the quantum case
the polarization direction is \emph{not known} prior to the transmission and
the classical method is not applicable. In this case one must in general
consider a system-bath Hamiltonian that is a linear combination of all $10$
possible independent bilinear terms: $\mathcal{A}=\{b_{1}^{\dagger
}b_{2},b_{2}^{\dagger }b_{1},(b_{1}^{\dagger })^{2},(b_{2}^{\dagger
})^{2},(b_{1})^{2},(b_{2})^{2}\}$, $\mathcal{B}=\{b_{1}b_{2},b_{1}^{\dagger
}b_{2}^{\dagger }\}$, $\mathcal{C}=\{b_{1}^{\dagger }b_{1},b_{2}^{\dagger
}b_{2}\}$ (the grouping will be clarified momentarily). It can be shown that all $10$ of these terms can be
eliminated by installing $18$ linear optical
devices that include beam-splitters in addition to phase-shifters,
i.e., in $16$ elementary steps (we combine beam-splitting and
phase-shifting into one step). This
result is based on Eq.~(\ref{eq:n-b}) and the following identities [that
follow directly from Eq.~(\ref{eq:n-b})] 
\begin{equation}
e^{i\phi \hat{n}}(b^{\dagger })^{2}e^{-i\phi \hat{n}}=e^{2i\phi }(b^{\dagger
})^{2},\quad e^{i\phi \hat{n}}(b)^{2}e^{-i\phi \hat{n}}=e^{-2i\phi }(b)^{2}.
\label{eq:n-b2}
\end{equation}%
The role of the beam-splitter is to eliminate the set of operators
$\mathcal{C}$; the beam-splitter is inserted after the first eight
steps. The $16$-step result can be considerably simplified in the realistic situation
wherein the two polarizations used to represent our qubit are degenerate. In
this case $\mathcal{C}$ becomes $b_{1}^{\dagger }b_{1}+b_{2}^{\dagger }b_{2}$%
, which generates an \emph{overall} phase and hence will not cause
decoherence. In this degenerate case, as we now show, we need only
phase-shifters to eliminate all contributions to decoherence. Let 
\begin{equation}
\Pi _{i}=e^{i\pi \hat{n}_{i}},\quad \Gamma =e^{i\pi (\hat{n}_{1}-\hat{n}%
_{2})/2},
\end{equation}%
i.e., a pair of phase-shifters. It follows immediately from Eqs.~(\ref%
{eq:n-b}),(\ref{eq:n-b2}) that%
\begin{equation*}
\Gamma ^{\dagger }\mathcal{A}\Gamma =\mathcal{A}
\end{equation*}%
while 
\begin{equation}
\Gamma ^{\dagger }\mathcal{B}\Gamma =\mathcal{B},\quad \Pi ^{\dagger }%
\mathcal{A}\Pi =\mathcal{A},\quad \Pi ^{\dagger }\mathcal{B}\Pi =\mathcal{B}
\end{equation}
[where $\Pi =\Pi _{1}\Pi _{2}$ was used above]. From these and the results
for the \textquotedblleft anti-linear-decoherence fiber\textquotedblright ,
the sequence $\Omega _{12}\equiv \lbrack 2,\Pi ,1,\Pi ]$ does not contain
any linear terms, but still contains all bilinear terms. Then, the sequence 
\begin{eqnarray}
\Omega _{1234} &\equiv &[\Omega _{34},\Gamma ^{\dagger },\Omega _{12},\Gamma
]  \notag \\
&=&[4,\Pi ,3,\Pi \Gamma ^{\dagger },2,\Pi ,1,\Pi \Gamma ]
\end{eqnarray}%
has, in four elementary phase-shifter steps, eliminated $H_{I}^{l}$ as well
as $\mathcal{A}$, and in particular the polarization-direction-changing
terms $b_{1}^{\dagger }b_{2}$ and $b_{2}^{\dagger }b_{1}$: at this point we
have a \emph{polarization-preserving fiber}. Note that the composite terms
can be combined into a single phase-shifter, i.e., 
\begin{eqnarray}
\Pi \Gamma ^{\dagger } &=&e^{i\pi (\hat{n}_{1}+3\hat{n}_{2})/2},  \notag \\
\Pi \Gamma &=&e^{i\pi (3\hat{n}_{1}+\hat{n}_{2})/2}.
\end{eqnarray}

The only remaining bilinear terms at this point are the counter-rotating
terms $\mathcal{B}=\{b_{1}b_{2},b_{1}^{\dagger }b_{2}^{\dagger }\}$, that
are typically neglected in the rotating wave approximation \cite{Cohen-Tannoudji}. To eliminate them nevertheless, we note that 
\begin{equation*}
\Pi _{1}\mathcal{B}\Pi _{1}=-\mathcal{B}.
\end{equation*}%
Therefore the sequence that eliminates \emph{all} linear and bilinear terms
for degenerate qubit states is 
\begin{eqnarray*}
[\Omega _{5678},\Pi _{1},\Omega _{1234},\Pi _{1}]= 
[8,\Pi ,7,\Pi \Gamma ^{\dagger },6,\Pi ,5,\Pi \Gamma \Pi _{1},4,\Pi
,3,\Pi \Gamma ^{\dagger },2,\Pi ,1,\Pi \Gamma \Pi _{1}],
\end{eqnarray*}%
which involves $8$ elementary phase-shifter steps (note that $\Pi \Gamma \Pi
_{1}=e^{i\pi (5\hat{n}_{1}+\hat{n}_{2})/2}$). At this point we have a fiber
that is completely free of both linear and bilinear decoherence-causing
terms for degenerate polarization qubits.

We can repeat the mixed-classical-quantum rough distance estimate above, by
simply assuming that now contributions to decoherence come only due to third
order in $l$: $l^{3}N=10^{-4}$. This leads to $N=5/\sqrt{20}\approx 1.2$,
and recalling that $8$ phase-shifters are needed per cancellation step, we
arrive at an upper-bound estimate of about $10$ phase-shifters per km, or
one phase-shifter every 100m. These phase-shifters must be introduced in
addition to the ones used above for cancellation of first order effects. We
have again assumed here that the fiber is straight; local curvature may
impose a lower upper bound.

\section{General decoherence elimination}

So far we have considered linear and bilinear photon terms in the
interaction Hamiltonian. The most general two-mode photon-related term in a
Hamiltonian is $b_{1}^{\dagger r}b_{1}^{s}b_{2}^{\dagger k}b_{2}^{l}$.
Provided $r\neq s$ and $k\neq l$ the identity%
\begin{eqnarray*}
e^{i(\alpha n_{1}+\beta n_{2})}b_{1}^{\dagger r}b_{1}^{s}b_{2}^{\dagger
k}b_{2}^{l}e^{-i(\alpha n_{1}+\beta n_{2})}
=e^{i[(r-s)\alpha +(k-l)\beta }b_{1}^{\dagger r}b_{1}^{s}b_{2}^{\dagger
k}b_{2}^{l}
\end{eqnarray*}%
shows that such a term can be eliminated using only phase shifters. For
example, when $r+s+k+l$ is an odd number, our considerations in the linear
case show that the term can be eliminated using the phase shifter $\Pi $,
while $b_{1}^{\dagger 2}b_{2}^{2}$ can be eliminated using $e^{-i\frac{\pi }{%
2}n_{1}}$. High-order terms with $r,s,k,l>1$ arise if one considers the
relativistic contribution, and they appear also in most of the effective
photon scattering theories. It should be clear that if such terms arise,
they can be reduced using additional phase-shifters, or beam-splitters in
the case $r=s$ and/or $k=l$, which arise due to terms containing photon
number operators.

\section{Connection to known leading loss mechanisms in optical fibers}

\label{leading}

As mentioned in Section \ref{noise} the leading loss mechanisms in optical
fibers are well characterized:\ UV absorption, Rayleigh backscattering, OH
absorption, and infrared absorption. It is useful to quickly review how
these processes arise, and then are treated by our method. Consider, for
example, the case of Rayleigh backscattering. We base our discussion on the
standard reference \cite{Loudon:book} (for a general description of
absorption see p.168; the cross section of Raleigh scattering is given on
pp. 371-373). The discussion starts \cite{Loudon:book}[Eq. (4.9.9)] from the 
\emph{dipole approximation} to our general photon-matter interaction
Hamiltonian, Eq.~(\ref{eq:Hl}):%
\begin{widetext}
\begin{eqnarray}
\widehat{H}_{ED} &=&ie\sum_{\mathbf{k}}\sum_{\lambda }\sum_{i,j}(\hbar
\omega _{k}/2\varepsilon _{0}V)^{1/2}\mathbf{e}_{\mathbf{k}\lambda }\cdot 
\mathbf{D}_{ij} \{\widehat{b}_{\mathbf{k}\lambda }\exp (i\mathbf{k}\cdot \mathbf{R}%
)-\widehat{b}_{\mathbf{k}\lambda }^{\dagger }\exp (-i\mathbf{k}\cdot \mathbf{%
R})\}\left\vert i\right\rangle \left\langle j\right\vert
\end{eqnarray}%
\end{widetext}
where $\left\vert i\right\rangle $ is the interacting charged particle
state, or the eigenstate of $H_{M}$, $\mathbf{R}$ is the atom
position, $V=L^{3}$ is the volume, $\mathbf{D}_{ij}=-e\left\langle
i\right\vert \sum \mathbf{r}_{\alpha }\left\vert j\right\rangle $ are
the
matrix elements of atomic dipole moment, and $\lambda $ is the polarization.  A
general scattering transition rate $\tau $ is \cite{Loudon:book}[Eq.
  (7.7.2)]:
\begin{eqnarray}
\frac{1}{\tau } &=&\sum_{f}\sum_{\mathbf{k}_{sc}}\left\vert \sum_{l}\frac{%
\left\langle n-1,1,f\right\vert \widehat{H}_{ED}\left\vert l\right\rangle
\left\langle l\right\vert \widehat{H}_{ED}\left\vert n,0,1\right\rangle }{%
n\omega -\omega _{l}}\right\vert ^{2} 
\frac{2\pi }{\hbar ^{4}}\delta (\omega _{f}+\omega _{sc}-\omega ).
\label{eq:tau}
\end{eqnarray}%
where $\left\vert 1\right\rangle $ and $\left\vert f\right\rangle $ are
the atomic ground state and final state. Initially, there are $n$ photons with
with frequency $\omega $ and wave vector $\mathbf{k}$. At the end there are $%
n-1$ incident photons and a single scattered photon with frequency $\omega
_{sc}$ and wave vector $\mathbf{k}_{sc}$.
Then the cross-section follows
from the relation $\sigma (\omega )=V/cn\tau $, and the differential
light-scattering cross-section is $\frac{d\sigma (\omega )}{d\Omega }$. The
differential cross-section of Rayleigh scattering is the special case when
the atom returns to its ground state, which is \cite{Loudon:book}[Eq.
(8.8.1)]:
\begin{eqnarray}
\frac{d\sigma (\omega )}{d\Omega }=\frac{e^{4}\omega ^{4}}{16\pi
^{2}\varepsilon _{0}^{2}\hbar ^{2}c^{4}}\left\vert \sum_{l}\frac{(\mathbf{e}
_{sc}\cdot \mathbf{D}_{1l})(\mathbf{e}\cdot \mathbf{D}_{l1})}{\omega
  _{l}-\omega } \right. 
+ \left. \frac{(\mathbf{e}\cdot \mathbf{D}_{1l})(\mathbf{e}_{sc}\cdot 
\mathbf{D}_{1l})}{\omega _{l}+\omega }\right\vert ^{2},
\end{eqnarray}
where the parameters are obtained from matrix elements of $\widehat{H}_{ED}.$

The important equation is (\ref{eq:tau}) above: it shows that Rayleigh
scattering originates from the interaction $\widehat{H}_{ED}$. Clearly, the
differential cross-section of Rayleigh scattering vanishes when $\widehat{H}%
_{ED}$ is zero. \emph{Our spatial BB\ method does just that: it effectively
eliminates the interaction} $\widehat{H}_{ED}$. Of course, this is not
unique to Rayleigh scattering, which is just one of the processes derived
from considering various cases involving $\widehat{H}_{ED}$. For example,
photon absorption and emission is mainly related to transitions involving
two atomic or molecular levels. The corresponding matrix element for
absorption is \cite{Loudon:book}[Eq. (4.10.1)]: 
\begin{eqnarray}
\left\langle n_{\mathbf{k}\lambda }-1, 2 \left\vert 
\widehat{H}_{ED}\right\vert n_{\mathbf{k}\lambda },1\right\rangle
=i\hbar g_{k\lambda } \exp [i(\omega _{0}-\omega
  _{k})t+i\mathbf{k\cdot R}]n_{\mathbf{k}\lambda }^{1/2},
\label{eq:absor}
\end{eqnarray}
where $g_{k\lambda }=(e\omega _{k}/2\varepsilon _{0}\hbar V)^{1/2}\mathbf{e}%
_{\mathbf{k}\lambda }\cdot \mathbf{D}_{12}$. The radiative lifetime is 
\begin{equation}
1/\tau _{R}=2\pi \sum_{\mathbf{k}}\sum_{\lambda }g_{\mathbf{k}\lambda
}^{2}\delta (\omega _{k}-\omega _{0}),
\end{equation}%
and, of course, it follows from Eq.~(\ref{eq:absor}) that this absorption is
prevented when $\widehat{H}_{ED}$ is zero.

Note how $\widehat{H}_{ED}$, which is effectively eliminated by our method,
involves the bosonic raising and lowering operators $\widehat{b}_{\mathbf{k}%
\lambda },\widehat{b}_{\mathbf{k}\lambda }^{\dagger }$. The reason that our
method is so general is that it acts directly on these operators, and
\textquotedblleft time-reverses\textquotedblright\ $\widehat{H}_{ED}$ by
flipping their sign.

\section{Proposal for an experiment}

As mentioned above, a crucial requirement for the success of our proposed
method is to insert the optical elements at intervals over which the
coherence-loss is still quadratic, rather than exponential. An experiment to
test for this regime is thus useful. This could be done by monitoring the
coherence (in particular, loss) locally, by focusing onto the edge of the
fiber and collecting light into a photon-counting device (since the absolute
intensity would be very small). By moving the focus along the fiber, one
should be able to track the decay as a function of distance from the fiber
entry point, and observe the required quadratic-to-exponential transition,
yielding an estimate of $\Delta $.

To actually test the method in the presence of phase-shifters, one could
repeat the above experiment with a single fiber and write some phase-shift
segments into it (as in the manufacturing of fiber Bragg gratings), at
intervals bounded above by those determined from the first experiment. We
note that a point of some potential concern is the impedance mismatch
between air and the phase-shifter material, that will lead to reflection.
Let $n_{i}$ ($i=1,2$) denote the indices of refraction: the reflected
amplitude is $(n_{2}-n_{1})/(n_{2}+n_{1})$, which leads, at normal
incidence, to 4\% loss per air-glass interface. However, a standard
anti-reflection coating can solve the problem: a quarter-wave layer of
material at $\sqrt{n_{1}n_{2}}$ between the two materials (two equal
reflections out of phase cancel out). In fibers the index changes will be
smaller and reflection is typically neglected. Moreover, by writing a smooth
phase profile as in the experiment proposed above, the reflection problem
essentially disappears.

Once $\Delta $ has been estimated, one can proceed to directly test our
method, as follows. Take two fiber segments and write a $\pi $ phase-shifter
(PS) into each. Attach them co-linearly (i) in the order PS-fiber-PS-fiber,
(ii) in the order fiber-PS-PS-fiber, and perform a photon counting
measurement. Our method should reduce attenuation in (i) by comparison to
(ii).

\section{Conclusions}

We have proposed a method to reduce quantum noise in optical fibers, via the
insertion of phase-shifters at appropriately spaced intervals. We have shown
that, in principle, this method can eliminate all quantum noise processes
that do not involve photon number operators in the system-bath Hamiltonian;
when such terms do arise, the phase shifters need to be supplemented with
beam-splitters, and our conclusions remain. Thus, with simple linear-optical
devices, quantum noise in optical fibers can be drastically reduced. This
conclusion has potentially important implications for quantum communication
(and its variants, quantum cryptography and distributed quantum computing)
via optical fibers. The practical feasibility of our method hinges on the
required distance between phase-shifters. We have given a rough upper-bound
estimate of several meters based on known attenuation rates. We have also
presented a more detailed calculation that predicts a range of distances,
depending on the bath spectral density appropriate for a fiber. Ultimately
we believe that the best way to test our proposal is to perform the
relatively straightforward experiment that it implies.

\begin{acknowledgments}
Support from NSERC, the DARPA-QuIST program (managed by AFOSR under
agreement No. F49620-01-1-0468), and the Sloan Foundation is gratefully
acknowledged (to D.A.L.). We thank Prof. H.-K. Lo, Prof. A.M. Steinberg, Prof.
T. Sargent, and Dr. Y.Z. Sun for very helpful
discussions.
\end{acknowledgments}


\appendix{}

\section{Detailed model for estimating $\Delta $}

\label{app}

Recall that our main approximation was the assumption of average fiber
homogeneity, Eq.~(\ref{eq:approxH}). In this appendix we relax this
assumption in order to estimate an upper bound on the distance $\Delta $
between phase shifters. We do this by considering corrections to order $\tau
^{2}$ and the non-ideal case 
\begin{eqnarray}
H_{0}((k-1)\Delta ) &=&H_{0}(k\Delta )+\varepsilon P_{k},  \notag \\
H_{I}^{l}((k-1)\Delta ) &=&H_{I}^{l}(k\Delta )+\varepsilon Q_{k},
\end{eqnarray}%
where $\varepsilon \ll 1$ and we take $P_{k},Q_{k}$ to be independent,
identically distributed (IID) Gaussian, local, and time-dependent
operator-valued corrections. This phenomenological model of fiber
inhomogeneity may be the result of material non-uniformity along the fiber
(such as local defects), slow time-dependent fluctuations in fiber
properties, or even the quadratic interaction~(\ref{eq:q}). By virtue of the
central limit theorem it will be accurate in the case of a \emph{large number%
} of defects. We assume that the effective BB time-interval $\tau $ is
chosen to be on the order of the small parameter $\varepsilon $ (though we
make no attempt to estimate $\varepsilon $). In this case, using the BCH
formula $e^{A}e^{B}=e^{A+B+[A,B]/2+...}$ to second order (i.e., keeping only
terms of order $\varepsilon ,\tau ,\varepsilon ^{2},\varepsilon \tau ,\tau
^{2}$), we find instead of the ideal Eq.~(\ref{eq:cancel}): 
\begin{widetext}
\begin{eqnarray}
e^{-iH((k-1)\Delta )\tau }\Pi e^{-iH(k\Delta )\tau }\Pi 
&=&e^{-iH((k-1)\Delta )\tau }e^{-i\Pi H(k\Delta )\Pi \tau }  \notag \\
&=&e^{-i[H_{0}(k\Delta )+H_{I}^{l}(k\Delta )+\varepsilon (P_{k}+Q_{k})]\tau
}e^{-i[H_{0}(k\Delta )-H_{I}^{l}(k\Delta )]\tau }  \notag \\
&\approx &\exp \{-i\tau \lbrack 2H_{0}(k\Delta )+\varepsilon
(P_{k}+Q_{k})]-\tau ^{2}[H_{I}^{l}(k\Delta ),H_{0}(k\Delta )]\},
\end{eqnarray}%
where in the second line the effect of the phase shifters was to flip the
sign (and thus cancel) the $H_{I}^{l}(k\Delta )$ term. To the same order of
accuracy the overall evolution operator becomes 
\begin{equation}
U^{\prime }(T,0)\approx e^{-iH_{0}(0)T}\exp \{-\tau
^{2}\sum_{k=1}^{N/2}[H_{I}^{l}(2k\Delta ),H_{0}(2k\Delta )]\}\exp
\{-i\varepsilon \tau \sum_{k=1}^{N/2}(P_{2k}+Q_{2k})\}.  \label{eq:U'}
\end{equation}%
Let us evaluate the first exponential. Using Eqs.~(\ref{eq:H0}),(\ref{eq2}): 
\begin{eqnarray}
-i[H_{I}^{l}(2k\Delta ),H_{0}(2k\Delta )] &=&-i\sum_{j,j^{\prime }}[(\hat{B}%
_{j}^{\dagger }(2k\Delta )b_{j}+\hat{B}_{j}(2k\Delta )b_{j}^{\dagger
}),\hbar \omega _{j^{\prime }}(2k\Delta )(\hat{n}_{j^{\prime
}}+1/2)+H_{M}(2k\Delta )]  \notag \\
&=&-i\sum_{j=0,1}\{\hbar \omega _{j}(2k\Delta )\hat{B}_{j}(2k\Delta )+[\hat{B%
  }_{j}(2k\Delta ),H_{M}(2k\Delta )]\}b_{j}^{\dagger } \notag \\
&&-\{\hbar \omega
_{j}(2k\Delta )\hat{B}_{j}^{\dagger }(2k\Delta )-[\hat{B}_{j}^{\dagger
}(2k\Delta ),H_{M}(2k\Delta )]\}b_{j}  \notag \\
&\equiv &H^{\prime },
\end{eqnarray}%
where $H^{\prime }$ is an effective Hamiltonian (it is Hermitian), which
plays the role of a Lamb shift \cite{Lidar:CP01}. We thus have for the first
exponential in Eq.~(\ref{eq:U'}): 
\begin{equation}
\exp \{-\tau ^{2}\sum_{k=1}^{N/2}[H_{I}^{l}(2k\Delta ),H_{0}(2k\Delta
)]\}=\exp (-i\tau ^{2}H^{\prime }),
\end{equation}%
whose effect is an energy renormalization (i.e., a phase shift), and does
not contribute to decoherence.

Next, consider the second exponential in Eq.~(\ref{eq:U'}). The operator $%
G(t)$ defined through $\sum_{k=1}^{N/2}(P_{2k}+Q_{2k})\sim
\int_{0}^{T}[P(t)+Q(t)]dt\equiv \int_{0}^{T}G(t)dt$ is Gaussian distributed
by our assumption that $P_{2k},Q_{2k}$ are Gaussian, IID random variables.
We would like to estimate the average deviation in $U^{\prime }(T,0)$ that
results from its presence. Since $G(t)$ is Gaussian distributed the average
can be computed as follows \cite{Leggett:87}:

\begin{eqnarray}
\langle \exp \{-i\varepsilon \tau \sum_{k=1}^{N/2}(P_{2k}+Q_{2k})\}\rangle 
&\sim &\langle \exp [-i\varepsilon \tau \int_{0}^{T}G(t)dt)]\rangle   \notag
\\
&=&\exp [-i\varepsilon \tau \int_{0}^{T}\int_{0}^{T}\langle G(t)G(t^{\prime
})\rangle dtdt^{\prime }]  \notag \\
&\equiv &\exp [-\varepsilon \tau \Gamma (T)].
\end{eqnarray}%
Expressed in terms of Fourier components $G_{\omega }$ of $G(t)$ we have for
the decoherence factor: 
\begin{equation}
\Gamma (T)=\frac{1}{2}\int_{0}^{\infty }d\omega \langle G_{\omega
}^{2}\rangle Q(\omega ,T)
\end{equation}%
where 
\begin{equation}
Q(\omega ,T)=\int_{0}^{T}\int_{0}^{T}dt\,dt^{\prime }\,\cos (\omega
(t-t^{\prime }))=\left( \frac{2\sin (\omega T/2)}{\omega }\right) ^{2}
\end{equation}%
But in the Gaussian case we have (as in the spin-boson model \cite%
{Leggett:87})%
\begin{equation}
\langle G_{\omega }^{2}\rangle =\frac{1}{2}I(\omega )\coth \frac{\beta
\omega }{2},
\end{equation}%
where $I(\omega )$ is the spectral density (of matter in the fiber) and $%
\beta $ is the inverse temperature. Hence our result is that the correction
is

\begin{equation}
\exp [-\varepsilon \tau \Gamma (T)]=\exp \left[ -\varepsilon \tau
\int_{0}^{\infty }d\omega I(\omega )\coth \frac{\beta \omega }{2}\left( 
\frac{\sin (\omega T/2)}{\omega }\right) ^{2}\right] .  \label{eq:SBint}
\end{equation}%
The attenuation is thus strongly dependent upon the form of $I(\omega )$,
but also depends sensitively on temperature. In particular, the thermal
time-scale $\hbar \beta $ is important in separating thermal effects from
effects due purely to vacuum fluctuations \cite{Palma:96}. In order to
formally separate the two it is convenient to write 
\begin{equation}
\coth \frac{\beta \omega }{2}=1+\bar{n}(\omega ,\beta )
\end{equation}%
where 
\begin{equation}
\bar{n}(\omega ,\beta )=\exp (-\beta \omega /2)/\sinh (\beta \omega /2)
\end{equation}%
is the average number of field excitations at inverse temperature $\beta $.

In the limit of very low temperatures ($\beta \gg 1$) we have 
\begin{equation}
\bar{n}(\omega ,\beta )\overset{\beta \gg 1}{\approx }2\exp (-\beta
\omega )  \label{eq:n-T-finite}
\end{equation}%
and we can analytically evaluate the integral in Eq.~(\ref{eq:SBint}), e.g.,
for the class of Ohmic-type spectral densities, i.e., for the case 
\begin{equation}
I(\omega )=\alpha \omega ^{n}e^{-\omega /\omega _{c}},
\end{equation}%
where $\alpha $ is the coupling strength and $\omega _{c}$ is the
high-frequency cutoff (note that $\alpha $ is not dimensionless). The result
in the zero-temperature case is 
\begin{eqnarray}
\lim_{\beta \rightarrow \infty }&\int_{0}^{\infty }&d\omega I(\omega )\coth 
\frac{\beta \omega }{2}\left( \frac{\sin (\omega T/2)}{\omega }\right)
^{2}= \notag \\
&&\left\{
\begin{array}{lr}
\frac{\alpha }{4}\log (1+(\omega _{c}T)^{2}), & \quad n=1 \\ 
\frac{\alpha }{2}\omega _{c}^{n-1}\Gamma (n-1)\left( 1-(1+(\omega
_{c}T)^{2})^{\frac{n-1}{2}}\cos [(n-1)\arctan (\omega _{c}T)]\right) ,
& \quad n\neq 1
\end{array}
\right. .
\end{eqnarray}
To obtain the non-zero temperature correction in the approximation (\ref%
{eq:n-T-finite}) take these results, multiply by $2$, replace $\omega _{c}$
by $\frac{\omega _{c}}{1+\beta \omega _{c}}$ everywhere, and add to the zero
temperature case. We tabulate a few cases of interest in the zero
temperature limit, letting $x\equiv \omega _{c}T$: 
\begin{equation}
\lim_{\beta \rightarrow \infty }\exp [-\varepsilon \tau \Gamma (T)]=\left\{ 
\begin{array}{lr}
(1+x^{2})^{-\alpha \varepsilon \tau /4}, & \quad n=1\text{ (Ohmic)} \\ 
\exp [-\frac{1}{2}\alpha \varepsilon \tau \omega _{c}\frac{x^{2}}{1+x^{2}}
], & \quad n=2\text{ (super-Ohmic)} \\ 
\exp [-\frac{1}{2}\alpha \varepsilon \tau \omega _{c}^{2}\frac{x^{2}(3+x^{2})
}{(1+x^{2})^{2}}], & \quad n=3\text{ (Debye)}
\end{array}%
\right. .
\end{equation}%
Let $1-\delta (T)$ be the desired coherence value after time $T$ (or
distance $X$); then we need to solve for the phase shifter spacing $\Delta $
from 
\begin{equation}
\lim_{\beta \rightarrow \infty }\exp [-\varepsilon \tau \Gamma (T)]>1-\delta
(T).
\end{equation}%
We find (assuming $\alpha >0$): 
\begin{equation}
\begin{array}{lr}
\Delta ^{2}<-4v^{2}\ln [1-\delta (T)]/\ln [(1+x^{2})], & \quad n=1 \\ 
\Delta ^{2}<-\frac{2v^{2}}{\alpha \omega _{c}}\frac{1+x^{2}}{x^{2}}\ln
[1-\delta (T)], & \quad n=2 \\ 
\Delta ^{2}<-\frac{2v^{2}}{\alpha \omega _{c}^{2}}\frac{(1+x^{2})^{2}}{%
x^{2}(3+x^{2})}\ln [1-\delta (T)], & \quad n=3%
\end{array}
.  \label{eq:Delta}
\end{equation}

The present model is, unfortunately, too phenomenological to make a reliable
estimate of $\Delta $. Nevertheless, it is of some interest to see its
prediction. E.g., we could wish to improve upon the current figure of merit
of $0.25$ db/km to the threshold value of $\delta (T)=10^{-4}$. Recall that $%
T=X/v$, $\tau =(\Delta /v)$ and we assumed $\tau \sim \varepsilon $. The
coupling strength $\alpha $ is typically of order unity \cite%
{Viola:98,Vitali:01}; we shall set $\alpha =1$. We take $v=c/1.6$, the speed
of light in a typical fiber, and $\delta (T)=10^{-4}$. The results in the
three cases, with $x=1.6/3\times 10^{-5}\omega _{c}$, are displayed in Fig.~%
\ref{fig}, as a function of the high-frequency cutoff $\omega _{c}$. As a
rough reference, the Debye temperature of amorphous silica is $T_{D}=342%
\mathrm{K}$ \cite{Freeman:86}, yielding a Debye frequency
estimate of $\omega _{c}=k_{B}T_{D}/\hbar =2\times 10^{13}\mathrm{Hz}$. The
corresponding value of $\Delta $ is $6\times 10^{5}\mathrm{m}$ ($n=1$), $0.6%
\mathrm{m}$ ($n=2$), $10^{-7}\mathrm{m}$ ($n=3$). This strong sensitivity to
the decoherence model underscores the need for the proposed experiment in
order to settle the question of the actual required distance between phase
shifters.

\begin{figure}[tbp]
  \includegraphics[height=8cm,angle=0]{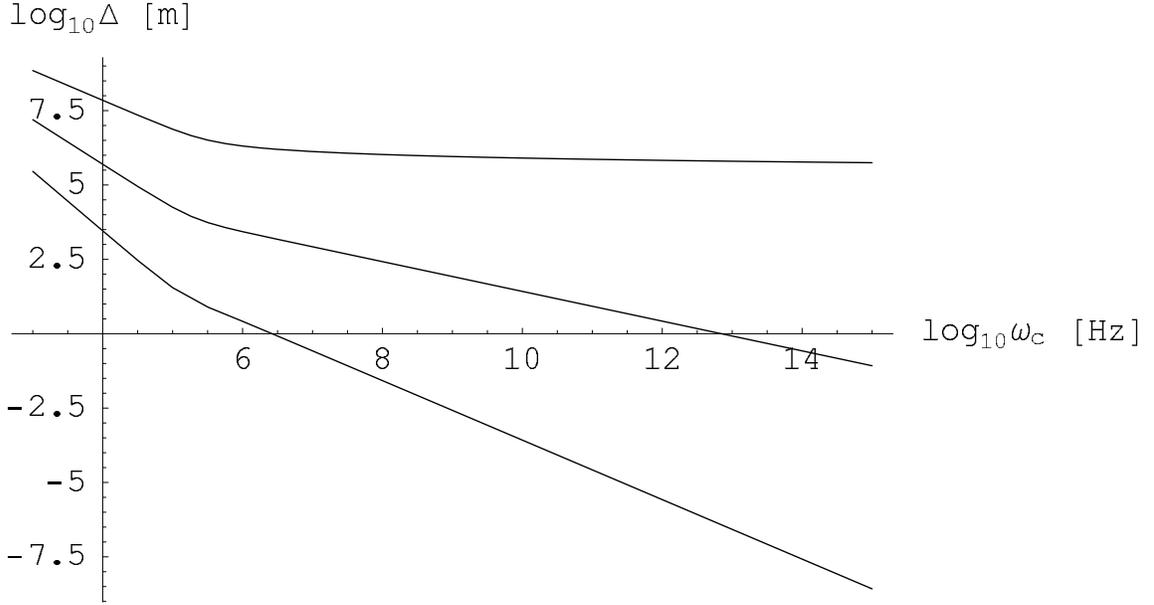}
\caption{Zero-temperature estimate of distance $\Delta $ between phase
shifters (in meters), as a function of high-frequency cutoff $\protect\omega %
_{c}$ (in Hz). Note the double logarithmic scale. Upper, middle, bottom
curves correspond to $n=1,2,3$ respectively in Eq.\emph{~}(\protect\ref
{eq:Delta}).}
\label{fig}
\end{figure}
\end{widetext}

\end{document}